\documentclass[12pt]{article} 

 \title{Interacting Bosons at Finite Temperature: \\
 How Bogolubov Visited a Black Hole and Came Home Again}

 \author{S. A. Fulling\\ 
\it Department of Mathematics and\\
\it Institute for Quantum Studies, Department of Physics,\\ 
\it Texas A\&M University, College Station, TX 77843-3368, USA\\
{\tt fulling@math.tamu.edu}
\and B.-G. Englert\\
\it Department of Mathematics and\\
\it Institute for Quantum Studies, Department of Physics,\\
\it Texas A\&M University, College Station, TX 77843-4242, USA;\\ 
\it Max-Planck-Institut f\"ur Quantenoptik,\\
\it Hans-Kopfermann-Strasse 1, 85748 Garching, Germany  
\and M. D. Pilloff\\ 
\it Institute for Quantum Studies, Department of Physics,\\
\it Texas A\&M University, College Station, TX 77843-4242, USA;\\ 
\it Department of Physics, University of California,\\ 
\it Berkeley, California, 94720, USA}

\date{July 5, 2002}

\def\onlinecite{\cite} 
\def\>{\rangle}  
\def\<{\langle}
\def\z#1#2{\ifcase#1{{\overline{#2}}}                
  \or\or\or\or\or\or\or\or\or 
  {{\mathbf{#2}}}\fi}                                 
\def\dg{\dagger}
\newcommand{\adj}{^{\dagger}}
\newcommand{\half}{\frac{1}{2}}

\begin{document} 
\maketitle 
 
\bigskip
\centerline{To appear in a special issue of \emph{Foundations of Physics}
in honor of Jacob Bekenstein}
\bigskip\goodbreak


 \begin{abstract}
 The structure of the thermal equilibrium state of a weakly interacting Bose gas 
is of current interest. 
We calculate the density matrix of that state in two ways.
The most effective method, in terms of yielding a simple, explicit answer,
is to construct a generating function within the traditional framework of quantum 
statistical mechanics. The alternative method, arguably more interesting, is to 
construct the thermal state as a vector state in an artificial system with twice 
as many degrees of freedom.  
It is well known that this construction has an actual physical realization in the 
quantum thermodynamics of black holes, where the added degrees of freedom 
correspond to the second sheet of the Kruskal manifold and the thermal vector 
state is a state of the Unruh or the Hartle--Hawking type.
What is unusual about the present work is that the Bogolubov transformation used 
to construct the thermal state  combines in a rather symmetrical way with 
Bogolubov's original 
transformation of the same form, used to implement the interaction of the nonideal 
gas in linear approximation.
 In addition to providing a density matrix, the method 
makes it possible to 
calculate efficiently certain expectation values  directly in terms of the thermal 
vector state of the doubled system.  
 \end{abstract}

 \section{Introduction}

 The temperature of black holes, predicted and investigated by 
Jacob Bekenstein on the basis of classical thermodynamic and 
statistical reasoning \cite{Bek,BekM}, 
 is intimately related to the mixing of normal modes in quantum 
field theory.
 In the work of Hawking \cite{Haw} (see also \onlinecite{otherBH})
 the disappearance of half the 
degrees of freedom of a field into the black hole was shown to be 
responsible for the entropy and temperature of the remaining 
degrees of freedom as an astrophysical black hole evaporates.
 Unruh \cite{Un} constructed radiating and 
(implicitly --- cf.\ \onlinecite{HH,Ful}) 
equilibrium states of a quantum field on the maximal 
Kruskal extension (``eternal'' black hole) in terms of coherent 
combinations of normal modes 
 on the two sheets of the wormhole.
 The analytic continuation involved in the Unruh construction 
 (which tells precisely how to combine positive-frequency modes on 
the physical sheet with negative-frequency modes on the unphysical 
sheet to create a state for the physical modes alone that is 
thermal at infinity)
 was soon recognized \cite{GP,Sew,revs} 
 to be an instance of the generic analyticity property of thermal 
states in a complex time coordinate \cite{KMS,HHW}. 
 Parker \cite{Par} showed that the well-known mixing of     positive-
 and negative-frequency modes in cosmological models would give 
rise to a thermal spectrum of created particles under some 
(restricted but plausible) conditions.
 (In Parker's scenario the two members of a correlated pair are 
both physical, but they are effectively rendered decoherent by 
spatial separation.) 
 
 Israel \cite{Is} and Sewell \cite{Sew} recognized Unruh's construction as a 
physical 
realization of a more abstract construction already known in 
quantum statistical mechanics 
 \cite{AW,HHW,TU}.
 In the latter, a fictitious copy of the physical system is 
introduced and its modes are mixed  with the physical modes to 
produce a thermal state of the latter.  (We review the details in 
the next section, for the case of a free boson system.)  
 In Unruh's scenario the second set of modes is not fictitious.
At worst, it resides on the second sheet of the Kruskal wormhole;
 for a uniformly
 accelerating observer in flat space-time
or an inertial observer in de Sitter space,
 it belongs to the 
part of ordinary space-time beyond the observer's horizon
 \cite{Un,BW,GH,Ful}.

 Black-hole temperature and cosmological particle production are 
effects in linear (noninteracting) quantum field theory in curved 
space-time that amount mathematically to linear mixing of creation 
and annihilation operators in a way that preserves their canonical 
commutation (or anticommutation) relations.
 Typically one set of operators defines a Fock-space structure so 
that the state under investigation is the corresponding vacuum, or 
no-quantum, state, while the other set of operators corresponds 
directly to physical observations.
 This type of construction 
(which pervades \onlinecite{Haw,otherBH,Un,Ful,Par})
is called a \emph{Bogolubov 
transformation}.

It is important to note that originally the Bogolubov transformation 
had nothing to do with either gravitational physics or temperature.
N.~N.~Bogolubov \cite{B,Bbook} introduced it to find the \emph{ground} 
state of a system of \emph{interacting} massive bosons, such as 
atoms.
 More precisely, the original application was to
ultracold helium,
 for which it turned out not to be quite adequate; but
nevertheless,  it has
become a central tool in condensed-matter physics. In recent
years, the necessity of a more detailed understanding of ultracold
gases of weakly interacting bosons brought about many new applications
of Bogolubov's formalism in  physical contexts where the underlying
approximations are quite reliable.
A recent article of Zagrebnov and Bru \cite{ZB} discusses these approximations, 
provides additional references to Bogolubov's pioneering work, and reviews the 
current mathematical status of the theory.

Our exposition of  Bogolubov's theory follows that of 
Kocharovsky et al.\ \cite{KKS}.
 Under the assumption of a two-body interaction, the Hamiltonian 
is of the form
 \begin{equation}
H = \sum_\z9k {\hbar^2 \z9k^2 \over 2M} a_\z9k^\dg a_\z9k
 + \sum_{\z9k_j} U_{\z9k_1\z9k_2,\z9k_3\z9k_4} 
a_{\z9k_1}^\dg a_{\z9k_2}^\dg a_{\z9k_3} a_{\z9k_4}\,.
 \label{Ham-true}\end{equation}
 If the interaction $U$ is sufficiently weak, most of the atoms 
will remain in  the ground mode of the noninteracting theory
 (``Bose condensate''), and the excitations out of that mode can 
be treated to good approximation by a linearization of~$H$.
One introduces new operators
 \begin{equation}
  \beta_0 = {a_0 \over \sqrt{a_0^\dg a_0 +1}}\,, \qquad
 \beta_\z9k= \beta_0^\dg a_\z9k\quad\hbox{for $\z9k\ne 0$}.
\label{beta-def} \end{equation}
 The operators with $\z9k\ne 0$ satisfy the canonical relations
 $ [\beta_\z9k,\beta_{{\z9k}'}^\dg] = \delta_{\z9k,\z9k'} $,
and one can assume that terms in $H$ containing more than two of 
them are negligible, so that $H$ is approximately of the form
 \begin{equation}
 H_B = E_0 + 
 \sum_{\z9k\ne0} V_\z9k \beta_\z9k^\dg \beta_\z9k
 + \sum_{\z9k\ne0} (W_\z9k  \beta_\z9k \beta_{-\z9k}
 + \hbox{H.c.}),
 \label{Ham-Bogo}\end{equation}
with coefficients independent of the sign of~$\z9k$.
 (Conservation of momentum and of parity have been assumed here.
 At the next step we assume positivity of the energy.)
 It is well known how to 
``diagonalize'' such a quadratic Hamiltonian by a Bogolubov 
transformation:
 \begin{equation}
 \beta_\z9k = u_\z9k b_\z9k + v_\z9k b_{-\z9k}^\dg,
 \qquad |u_\z9k|^2 - |v_\z9k|^2 =1;
 \label{Bogo-gen}\end{equation}
 \begin{equation}
 H_B = E_0 + \sum_{\z9k\ne0} \epsilon_\z9k b_\z9k^\dg b_\z9k.
\label{Ham-diag} \end{equation}
(All the undefined quantities in (\ref{Ham-Bogo}), 
 (\ref{Bogo-gen}), (\ref{Ham-diag}) can be calculated from the
 $U$ and $M$ in (\ref{Ham-true}); the details are not 
important here.)
 The dynamics of the system in the $b$ variables is now trivial;
 for physical interpretation one returns to the $\beta$ variables,
 for example, to see the structure of the physical ground state in 
terms of real atoms.

 In the wake of the recent experimental observations of Bose 
condensation of atoms \cite{expt}, interest in the type of problem 
originally investigated by Bogolubov has intensified.
 Methods have been invented to measure
directly the (magnitudes of the) amplitudes $u_{\z9{k}}$ and
$v_{\z9{k}}$ in (\ref{Bogo-gen}) in a Bose--Einstein condensate 
\cite{expt-amps}. 
 In \onlinecite{KKS} the (weakly) interacting Bose gas was studied at (low 
but) finite temperature;
 formulas were found for the characteristic function 
 (Fourier-transformed probability distribution) of the total number 
of atoms in a pair of modes, 
 $\beta_\z9k^\dg \beta_\z9k +  \beta_{-\z9k}^\dg \beta_{-\z9k}\,,$
 and some related statistical quantities
 (the generating cumulants).
 In \onlinecite{KKS} the interaction was treated as above, and the 
temperature was treated by constructing the statistical operator 
 (density matrix) for each 
mode in the traditional, direct way:
 \begin{equation}
 \rho_\z9k =(1-e^{-\epsilon_\z9k/T})
  e^{-\epsilon_\z9k b_\z9k^\dg b_\z9k/T}.
 \label{rho}\end{equation}
 (Note that it is the dressed ($b$) quanta that appear in 
(\ref{rho}), because the Hamiltonian is (\ref{Ham-diag}),
 but it is the bare ($\beta$) quanta whose statistics are being 
studied.)
 The formulas of \onlinecite{KKS} are expressed in terms of a quantity
 \begin{equation}
 z(A_\z9k) = {A_\z9k - e^{\epsilon_\z9k/T} \over
A_\z9k  e^{\epsilon_\z9k/T} -1}
 \label{A-def}\end{equation}
and $z(-A_\z9k)$,
  where $A_\z9k = v_\z9k/ u_\z9k $ is the parameter defining the 
Bogolubov transformation~(\ref{Bogo-gen}).

We recall that when $u_\z9k$ and $v_\z9k$ are real-valued,
  which one may assure by choosing the phases of the modes properly, 
 it is customary to write the coefficients in (\ref{Bogo-gen}) as 
 \begin{equation}
u_\z9k = \cosh \theta_\z9k, \quad
v_\z9k = \sinh \theta_\z9k; \qquad
 A_\z9k = \tanh \theta_\z9k\,.
\label{Bogo-hyp}\end{equation}
 (Henceforth we drop the subscript $\z9k$ whenever no ambiguity
results.)
 Then $\theta$ is an additive quantity precisely analogous to the 
rapidity parameter of a Lorentz transformation; 
 that is, the result of successively applying two Bogolubov 
transformations is the Bogolubov transformation corresponding to 
the sum of the respective $\theta$s.
 Now (\ref{A-def}) superficially resembles the addition formula for 
the hyperbolic tangent function,
 \begin{equation}
 \tanh (x\pm y) = 
 {\tanh x \pm \tanh y \over 1 \pm \tanh x \tanh y}\,.
 \label{tanh}\end{equation}
This suggests that the results of \onlinecite{KKS} can be better 
understood, and perhaps more easily derived, in terms of the 
composition of the transformation of the original Bogolubov type 
 (implementing the interaction) with a Bogolubov transformation of 
the Unruh type to implement the thermalization.
 Unfortunately, there are three things wrong with this naive idea.
 First, $e^{\epsilon/T}$ is greater than unity, and hence cannot be 
a hyperbolic tangent.
 Fortunately, that problem is instantly solved by rewriting 
(\ref{A-def}) as
 \begin{equation}
 -\,\frac1{z(A)} =
 {A  e^{\epsilon/T} -1 \over
 e^{\epsilon/T}- A} =
 {A -  e^{-\epsilon/T} \over
 1- A  e^{-\epsilon/T}}\,,
 \label{A-inv}\end{equation}
 which really does have the form (\ref{tanh}).
 Second [see (\ref{Bogo-T})], the correct Bogolubov parameter in 
the Unruh 
 (more properly, the Araki--Woods--Takahashi--Umezawa)
 construction is $e^{-\epsilon/2T}$, not $e^{-\epsilon/T}$.
 And third, the two Bogolubov transformations are about ``different 
axes'' in a space of dimension greater than~$2$,
 so there is no reason to expect them to compose in such a simple 
way.

 The present authors \cite{EFP} rederived (and generalized) the 
results of \onlinecite{KKS} in a simpler way, without, however, 
 elucidating any connection with the thermal Bogolubov 
construction.
 Here we present (in Secs.\ \ref{sec:2} and~\ref{sec:4}) a direct assault on  
the problem by the thermal Bogolubov method.
 Inevitably, the treatment displays a certain ``hidden symmetry'' 
between the parameters
 \begin{equation}
 \theta = \tanh^{-1} A \quad\hbox{and}\quad
 \phi = \tanh^{-1} e^{-\epsilon/2T}.
 \label{tanh-pars}\end{equation}
 Equally inevitably, this symmetry is largely destroyed in the 
final formula for the density matrix,
  since to get it one must trace over the 
fictitious modes of the thermal construction but does not trace 
over any of the physical modes involved in the Bogolubov dressing 
transformation (see Sec.~\ref{sec:3}).
 However, the symmetry is restored when one takes the expectation
 value of an observable that involves only \emph{one} physical 
traveling-wave mode.
In fact, the moments and cumulants associated with the number operators 
for one or two modes can be most easily calculated directly from the 
Araki--Takahashi--Unruh pure state vector, rather than from the density matrix
(Sec.~\ref{sec:5}).
In Sec.~\ref{sec:6}, to obtain a simpler form for the density-matrix elements 
(generalizing~\onlinecite{EFP})
we employ a different method, based on generating functions.

 \section{Main calculation}\label{sec:2}

 The analysis proceeds in seven steps. 

 \subsection{Reduction to standing-wave modes}\label{sec:2a}
 Introduce (for each pair $\pm\z9k$)
 \begin{equation}
 b_c = {b_\z9k + b_{-\z9k}\over \sqrt2}\,, \quad
b_s = {b_\z9k - b_{-\z9k}\over \sqrt2}\,,
 \label{cos-sin}\end{equation}
 so that 
\begin{eqnarray}
b_\z9k e^{i\z9k\cdot \z9x} + b_{-\z9k} e^{-i\z9k\cdot \z9x}
& =& \sqrt2 \bigl[b_c \cos {(\z9k\!\cdot\! \z9x})
   +ib_s \sin {(\z9k\!\cdot\! \z9x)}\bigr]\,,
\nonumber\\
 b_\z9k^\dg b_\z9k +b_{-\z9k}^\dg b_{-\z9k} &=& 
 b_c^\dg b_c + b_s^\dg b_s\,, 
\end{eqnarray}
 and these operators and their adjoints satisfy the canonical 
commutation relations.
 Henceforth we concentrate largely on the cosine modes and often 
omit the subscript $c$. 

 \subsection{Dressing by the interaction}\label{sec:2b}
The Bogolubov transformation (\ref{Bogo-gen}) and (\ref{Bogo-hyp})
is chosen to convert (\ref{Ham-Bogo}) into (\ref{Ham-diag}).
The inverse of such a transformation is the same except for the 
sign of $\theta$, $A$, and $v$.
 The transformation thereby induced on the cosine modes is
 \begin{equation}
 \beta = \cosh {\theta}\, b + \sinh {\theta}\, b^\dg; \qquad
  b = \cosh{\theta}\,\beta - \sinh{\theta}\,\beta^\dg; \qquad
 \theta = \tanh^{-1}A.
\label{Bogo-A} \end{equation}
The transformation of the sine modes has the opposite sense:
\begin{equation}
 \beta_s = \cosh {\theta}\, b_s - \sinh {\theta} \,b_s^\dg; \qquad
  b_s = \cosh{\theta}\,\beta_s + \sinh{\theta}\,\beta_s^\dg.
\label{Bogo-sin} \end{equation}
 (This strange sign 
 is an artifact of our phase convention in (\ref{cos-sin});
 it could be avoided at the cost of unnecessary imaginary numbers 
elsewhere.) 

 Recall that $\beta$ annihilates bare quasiparticles, here 
interpreted as excited physical atoms, while $b$ annihilates dressed 
quanta, so that the eigenstates of energy are eigenstates of the 
number operator $b^\dg b$.
 If $|0)$ is the ground state: $b|0) =0$, and $|n\>$ is the state of 
$n$ physical atoms: $\beta^\dg \beta |n\> = n|n\>$, then
 \begin{equation}
 |0) = \frac1{\sqrt{\cosh\theta}} \sum_{j=0}^\infty A^j 
 \sqrt{(2j-1)!! \over (2j)!!} \,|2j\>
 \qquad (A = \tanh \theta).
\label{vac-A} \end{equation}
 It is instructive to review the proof of (\ref{vac-A}):
 Write $|0)$ as $N \sum_{n=0}^\infty c_n |n\>$ with $c_0=1$.
 Impose the condition that 
 $(\cosh\theta \,\beta - \sinh {\theta}\, \beta^\dg)|0) = 0$
to obtain the recursion 
 $c_{n+1} =  \sqrt{n\over n+1}\tanh{\theta}\, c_{n-1}\,$.
 It follows that $c_n=0$ for odd $n$ and
 $c_{2j} = A^j \sqrt{(2j-1)!!/(2j)!!}$
 \thinspace(where $n!! = n(n-2) \cdots$).
Impose the requirement that $(0|0)=1$ and observe that
\begin{equation}
\frac1{\sqrt{1-x}} = 
 \sum_{j=0}^\infty x^j \,{(2j-1)!! \over (2j)!!} 
\end{equation}
to conclude that 
$N^2 = \sqrt{1-\tanh^2 \theta} = (\cosh\theta)^{-1}$.

 \subsection{Thermal dressing}\label{sec:2c}
The thermal Bogolubov construction must be applied to the energy 
eigenstates, regardless of whether there is an interaction.
So (working with one mode at a time) 
 we start from the number eigenstates satisfying
 \begin{equation}
b|0)=0\quad\mbox{and}\quad b^\dg b|n)= n|n)\,.  
 \end{equation}
Now introduce a fictitious mode ``on the other side of the world''
 with corresponding operators $\z0b$ and $\z0b^\dg$;
 we work in a doubled Fock space with basis vectors satisfying 
 \begin{equation}
 b|0,0)= 0\,, \quad \z0b|0,0)=0\,;\qquad
 b^\dg b|n,m)= n|n,m)\,, \quad \z0b^\dg \z0b |n,m) = m|n,m)\,.
 \end{equation}
The thermal Bogolubov transformation is \cite{TU,Un}
 \begin{equation}
c  = \cosh {\phi}\, b - \sinh {\phi}\, \z0b^\dg, \quad
  \z0c = \cosh{\phi}\,\z0b - \sinh{\phi}\,b^\dg; \qquad
e^{-\epsilon/2T} = \tanh \phi,
\label{Bogo-T} \end{equation}
with inverse 
 $ b= \cosh\phi \, c + \sinh\phi\, \z0c^\dg$,
$ \z0b = \cosh \phi\, \z0c + \sinh\phi\, c^\dg$.
 
 Let $|0,0\}$ be the state with no $c$ and $\z0c$ quanta:
 $c|0,0\}=0=\z0c|0,0\}$.
 Then
 \begin{equation}
|0,0\} = \frac1{\cosh\phi} \sum_{n=0}^\infty \tanh^n\phi\,|n,n) =
 (1-e^{-\epsilon/T})^{1/2}\sum_{n=0}^\infty e^{-n\epsilon/2t}
|n,n).
\label{vac-T} \end{equation}
 The proof is parallel to that of (\ref{vac-A}) and slightly 
simpler, because the alter\-nating-factorial and consequent 
 square-root functions do not arise.

 The statistical (density) operator for the whole system 
 (of one real and one fictitious mode) is
 \begin{equation}
|0,0\}\{0,0| = (1-e^{-\epsilon/T})\sum_{n,m=0}^\infty
 e^{-(n+m)\epsilon/2T} |n,n)(m,m|.
 \label{dens-full}\end{equation}
 The statistical operator for the physical mode alone is obtained by 
tracing over the states of the unphysical mode:
 \begin{equation}
\rho= (1-e^{-\epsilon/T})\sum_{n=0}^\infty
 e^{-n\epsilon/T} |n)(n|.
 \label{dens-phys}\end{equation}
 By design, $\rho$ is precisely [cf.~(\ref{rho})]
 the thermal ensemble at temperature~$T$!
 This is the generic Araki--Takahashi--Unruh construction, in the 
setting of our particular problem.

The foregoing notation is appropriate for the cosine modes.
 For the sine modes, we choose to change the sign of $\phi$
 in (\ref{Bogo-T}).  This convention corresponds to the natural,
 momentum-conserving Bogolubov transformation 
 (\ref{Bogo-Tk}) on the original 
traveling-wave modes (where $b_\z9k$ mixes with $b_{-\z9k}^\dg\,$,
 not $b_\z9k^\dg$),
 which is mandatory in an Unruh-type situation but arbitrary when 
the barred modes are completely fictitious.
The sign cancels in the final reduced density 
 matrix,~(\ref{dens-phys}). 

 \subsection{Diagonalization of the composite transformation}\label{sec:2d}
 We now combine (\ref{Bogo-A}) and (\ref{Bogo-T}) to express the 
bare modes in terms of the doubly dressed modes:
\begin{equation}
 \left(\begin{array}{l}
  \beta\\\beta^\dg\\\z0\beta\\\z0\beta^\dg \end{array}\right) =
 \left(\begin{array}{cccc}
 up&vp&vq&uq\\vp&up&uq&vq\\vq&uq&up&vp\\uq&vq&vp&up 
\end{array}\right)
 \left(\begin{array}{l}
 c\\c^\dg\\\z0c\\\z0c^\dg \end{array}\right).
 \label{matrix}\end{equation}
 (Here $p=\cosh\phi$, $q=\sinh\phi$, and $\z0{\beta}$ is related to 
$\z0b$ just as $\beta$ is to~$b$.)
Introduce
 \begin{equation}
 \Omega = \theta+\phi, \quad \Psi = \theta-\phi,
 \label{newvars}\end{equation}
 and
 \begin{equation}
 G = {\beta+\z0\beta \over\sqrt2}, \quad
H = {\beta-\z0\beta \over\sqrt2}, \qquad
 \gamma = {c+\z0c \over\sqrt2}, \quad
\delta = {c-\z0c \over\sqrt2}. 
 \label{newops}\end{equation}
 Then a short calculation shows that
 \begin{equation}
G = \cosh\Omega \, \gamma + \sinh \Omega \, \gamma^\dg, 
 \quad
  H = \cosh\Psi \, \delta + \sinh \Psi \, \delta^\dg.
 \label{Bogo-diag}\end{equation} 
That is, (\ref{matrix}) decouples into two elementary Bogolubov 
transformations!
 Moreover, the  idea that  (\ref{tanh}) should be playing a role is 
vindicated by (\ref{newvars}).

 \subsection{Construction of the thermal state with interaction}\label{sec:2e}
Note that the state annihilated by $\gamma$ and $\delta$ is the 
same as $|0,0\}$, the one annihilated by $c$ and $\z0c$;
 after tracing over the barred mode, it will give us the thermal state 
we want.
 But to interpret that state we need to express it in terms of the 
 $\beta^\dg\beta$ number observable.

 The first step is to construct $|0,0\}$ within the $(G,H)$ Fock 
space.
This is done by applying the mathematics of (\ref{vac-A}) to the 
two transformations (\ref{Bogo-diag}) independently.
We need to define yet another basis:
 \begin{equation}
 G^\dg G |J,L] = J |J,L], \quad
  H^\dg H |J,L] = L |J,L].
 \label{GH-basis} \end{equation}
 Then
 \begin{equation}
 |0,0\} = \frac1{\sqrt{\cosh\Omega\cosh\Psi}} \sum_{j,l=0}^\infty
 \tanh^j\Omega  \tanh^l\Psi 
 \sqrt{(2j-1)!!\,(2l-1)!!\over (2j)!!\,(2l)!!} \,
 |2j,2l].
\label{vac-temp} \end{equation}

 The basis of physical interest is defined by
 \begin{equation}
 \beta^\dg\beta |n,m\> = n|n,m\>, \quad
\z0\beta^\dg\z0\beta |n,m\> = m|n,m\>.
\label{beta-basis}\end{equation}
In view of (\ref{newops}), the connection between $|J,L]$
and $|n,m\>$ is just a unitary transformation, introducing no 
further pair creation. 
 (Nevertheless, it is the source of most of the combinatorial 
complexity of our result.)
One has
 \begin{eqnarray}
 |2j,2l] &=& \frac1{\sqrt{(2j)!}} (G^\dg)^{2j} 
\frac1{\sqrt{(2l)!}} 
(H^\dg)^{2l}|0,0]  \nonumber\\
 &=& \frac1{2^{j+l}\sqrt{(2j)!\,(2l)!}} 
 \sum_{\kappa=0}^{2j} \sum_{\lambda=0}^{2l} 
{2j\choose \kappa}{2l\choose\lambda}(-1)^\lambda \nonumber\\
 &&{}\times
 (\beta^\dg)^{2j-\kappa+2l-\lambda} (\z0\beta^\dg)^{\kappa+\lambda}
|0,0\> \nonumber\\
 &=& \frac1{2^{j+l}} 
 \sum_{\kappa,\lambda} {(-1)^\lambda \sqrt{(2j)!\,(2l)!} \over \kappa!\, 
(2j-\kappa)!\, \lambda!\, (2l-\lambda)!}  \nonumber\\
 &&{}\times 
 \sqrt{(2j+2l-\kappa-\lambda)!\,(\kappa+\lambda)!}
 |2j+2l-\kappa-\lambda, \kappa+\lambda\>.
 \label{rot}\end{eqnarray}
 Introduce $m=\kappa+\lambda$ and define
 \begin{equation}
 Q(m,j,l) = \sum_{\lambda=\max(0,m-2j)}^{\min(m,2l)}
 {(-1)^\lambda \over \lambda!\, (2l-\lambda)!\, (m-\lambda)! \, 
(2j-m+\lambda)!}\,.
 \label{Q}\end{equation}
 Then
 \begin{equation}
|2j,2l] = {\sqrt{(2j)!\, (2l)!} \over 2^{j+l}}
 \sum_{m=0}^{2j+2l} \sqrt{(2j+2l-m)!\, m!} \,Q(m,j,l)\,
 |2j+2l-m, m\>.
 \label{rotated}\end{equation}

 Substituting (\ref{rotated}) into (\ref{vac-temp}) and simplifying,
 one obtains (for a cosine mode)
 \begin{eqnarray}
|0,0\}&=& \frac1{\sqrt{\cosh\Omega\cosh\Psi}}
 \sum_{\scriptstyle n,m=0 \atop \scriptstyle n+m \mathrm{\ even}}^\infty
  |n,m\> \sqrt{n!\,m!} \nonumber\\
&&{}\times \sum_{l=0}^p \tanh^{p-l} \Omega\, \tanh^l \Psi
\, {(2l)! \,(2p-2l)! \over 4^p\, l!\, (p-l)!}\, Q(m,p-l,l),
 \label{vac}\end{eqnarray}
 where $p = \frac12(n+m)$.
For a sine mode, a factor $(-1)^p$ should be inserted.

 \subsection{Evaluation of $Q$}\label{sec:2f}
 First note that whenever $\lambda$ is outside the range specified 
in (\ref{Q}), the summand is $0$ because at least one of the 
denominator factors is at  a pole of the gamma function.
 Therefore, one may extend the summation over 
 $-\infty<\lambda<\infty$, and it is not necessary to write the 
limits at all \cite{Knu}.

 Now define
 \begin{equation}
 P(a,c,m) = 2^{-m} \sum_\lambda (-1)^\lambda {c-a\choose m-\lambda}
 {a\choose\lambda}.
 \label{P}\end{equation}
 Then, on the one hand,
 \begin{equation}
 P(a,c,m) = P_m^{(c-a-m, a-m)}(0),
 \label{Jacobi}\end{equation}
 where $P_m^{(\alpha,\beta)}(x)$ is a Jacobi polynomial \cite{GR}.
 (This special value of the Jacobi polynomial apparently cannot be 
reduced to anything simpler; the very systematic software 
associated with \onlinecite{PWZ} identifies it as a certain hypergeometric 
function (also in \onlinecite{GR}) but nothing less.)
 On the other hand, we have
 \begin{equation}
 4^{-p}\, (2l)!\,(2p-2l)!\, Q(m,p-l,l) =
 2^{-n} P(2l,2p,m) \qquad \bigl(p = \textstyle\frac12(n+m)\bigr).
 \label{Q-P}\end{equation}
 In this notation (\ref{vac}) becomes
\begin{eqnarray}
|0,0\}&=& \frac1{\sqrt{\cosh\Omega\cosh\Psi}}
 \sum_{\scriptstyle n,m=0 \atop \scriptstyle n+m \mathrm{\ even}}^\infty
  |n,m\> \sqrt{n!\,m!} \nonumber\\
&&{}\times \sum_{l=0}^p \tanh^{p-l} \Omega\, \tanh^l \Psi
\, {1 \over 2^n\, l!\, (p-l)!}\, P(2l,2p,m).
 \label{vac-P}\end{eqnarray}

 \subsection{The density matrix}\label{sec:2g}
 The density matrix $\rho_{nn'}=\<n|\rho|n'\>$
 for a cosine mode is obtained by tracing 
$|0,0\}\{0,0|$ over the states of the fictitious partner mode. 
 That is, in a linear combination of objects $|n,m\>\<n',m'|$ one 
must set $m'=m$ and sum over $m$.
 Since $n-m$ and $n'-m'$ are constrained to be even, the result is 
$0$ unless $n-n'$ is even. One gets
 \begin{eqnarray}
\rho_{nn'}&=&0 \quad \hbox{if $n-n'$ is odd}, \label{odd}\\
\rho_{nn'}&=&\frac1{\cosh\Omega\cosh\Psi} \, 
 {\sqrt{n!\,n'!} \over 2^{n+n'}} \sum_{\scriptstyle m =0 \atop
 \scriptstyle n-m \mathrm{\ even}}^\infty m! \nonumber \\
 &&{}\times \sum_{l=0}^p \sum_{l'=0}^{p'}
{P(2l,2p,m) P(2l',2p',m) \over l!\,l'!\, (p-l)!\, (p'-l')!}\,
 \tanh^{p+p'-l-l'}\Omega \, \tanh^{l+l'}\Psi, \nonumber\\
 && p = {n+m\over2}, \quad p'={n'+m\over 2}, \quad n-n' 
 \mathrm{\ even}.
 \label{density}\end{eqnarray}
 (The limits on the $l$ and $l'$ summations are superfluous, for 
the same reason explained earlier for~$\lambda$.)

 The density matrix for a sine mode is the same except for a factor
 $(-1)^{(n-n')/2}$.
 The density matrix for the entire atomic system is the tensor 
product of all the density matrices for the individual modes.
 (Recall that the latter depend on $|\z9k|$ through $\Omega$ 
and~$\Psi$.) 

 \section{Symmetries (or not)}\label{sec:3}
The \emph{Jacobi number} (\ref{P}) has these symmetries:
 \begin{eqnarray}
& P(c-a,c,m)= (-1)^m  P(a,c,m)\,;&
\nonumber\\
&2^m P(a,c,m) = (-1)^a 2^n P(a,c,n) \quad
 \hbox{if $n+m= c$}.&
\end{eqnarray}
 The first of these (with $c=2p$, $a=2l$) expresses the essential 
invariance of our formulas when a summation index $l$ is changed to  
$p-l$, 
 hence when $\Omega$ interchanged with $\Psi$.
 The second (with $a$ even) expresses the symmetry of $|0,0\}$ in 
the real and fictitious modes. 

 It is time to contemplate the meaning of $\Omega$ and $\Psi$ in 
terms of the basic parameters of our problem, $A$ and $\epsilon/T$.
    From the definitions we have
\begin{equation}
 \tanh\Omega= {A+e^{-\epsilon/2T} \over 1+ Ae^{-\epsilon/2T}}\,,
 \quad
 \tanh\Psi = {A-e^{-\epsilon/2T} \over 1- Ae^{-\epsilon/2T}}\,.
 \label{params}\end{equation}
 These bear tantalizing resemblances to $\frac1{z(-A)}$ and 
 $-\frac1{z(A)}$ as expressed through (\ref{A-inv}), but the factor 
of $2$ with the temperature is ineradicable.

 Now observe the behavior of the density matrix (\ref{density}) 
under elementary operations on the parameters:
 \begin{itemize}
 \item Interchanging $\Omega$ and $\Psi$
  (i.e., changing the sign of $\phi$) leaves $\rho$ invariant.
 Since this operation amounts to replacing $e^{-\epsilon/2T}$ by a 
negative number, it may appear unphysical.
 However, as previously remarked in connection with the sine modes,
 it really represents just an arbitrariness in the definition of 
the fictitious modes.
 \item Changing the signs of both $\Omega$ and $\Psi$ is equivalent 
to changing the signs of both $\theta$ and $\phi$.  Its effect on 
$\rho_{nn'}$ is an overall factor $(-1)^{(n-n')/2}$.
 This is precisely the distinction between cosine and sine modes.
\item Interchanging $\theta$ and $\phi$ is equivalent to changing 
the sign of $\Psi$.
 Its effect on $\rho$ is substantive:
 Each term in the summand is multiplied by $(-1)^{l+l'}$.
 Thus (of course) the final formulas of the theory are \emph{not} 
all symmetric under interchange of an interaction parameter with a 
temperature parameter, despite the intriguing symmetries in the 
formalism. 
 (We shall see later, however, that some formulas \emph{are} 
symmetric.)
 \end{itemize}
 The other $4$ nontrivial elements of the group generated by these 
operations add no additional insight.

 \section{A pair of modes}\label{sec:4}

At the cost of dealing with twice as many indices, 
 one can work directly with the original traveling-wave modes,
skipping step~(\ref{cos-sin}).
 Some of the intermediate results are useful, so we summarize the 
calculation here.

 Define a momentum-conserving thermal Bogolubov transformation
 \begin{equation}
 b_\z9k = p_\z9k c_\z9k + q_\z9k \z0c_{-\z9k}^\dg\,,
 \quad  \z0b_\z9k = p_\z9k \z0c_\z9k + q_\z9k c_{-\z9k}^\dg
 \label{Bogo-Tk}\end{equation}
equivalent to (\ref{Bogo-T}) and its $b_s$ counterpart.
When (\ref{Bogo-Tk}) is combined with (\ref{Bogo-gen}) and its barred 
counterpart, one obtains two $4\times4$ systems of precisely the 
form (\ref{matrix}), except that all the creation operators have 
the opposite sign of $\z9k$ from the annihilation operators they 
mix with.
So one can define operators as in (\ref{newops}), with subscripts 
$\pm$, and get
 \begin{equation}
 G_\pm = \cosh\Omega\, \gamma_\pm + \sinh\Omega\, \gamma_\mp^\dg\,,
 \quad
 H_\pm = \cosh\Psi\, \delta_\pm + \sinh\Psi\, \delta_\mp^\dg\,.
 \label{Bogo-diagk}\end{equation}

 By the same methods as before, one finds that the analog of 
(\ref{vac-temp}) is
 \begin{equation}
 |0,0,0,0\} =\frac1{\cosh\Omega\,\cosh\Psi}
 \sum_{j,l=0}^\infty
 \tanh^j\Omega  \tanh^l\Psi |j,j]\otimes|l,l],
\label{vac-tempk} \end{equation}
where $|0,0,0,0\}$ is annihilated by $\gamma_\pm$ and $\delta_\pm\,$, 
and
 \begin{eqnarray}
 G_+^\dg G_+ |j,\tilde \jmath] = j|j,\tilde \jmath], &\ &
H_+^\dg H_+ |l,\tilde l] = l|l,\tilde l], \nonumber\\
G_-^\dg G_- |j,\tilde\jmath] = \tilde\jmath|j,\tilde\jmath],  &\ &
H_-^\dg H_- |l,\tilde l] = \tilde l|l,\tilde l].
 \label{GH-basisk}\end{eqnarray}
 Note that in (\ref{vac-tempk}) the occupation numbers may be 
either even or odd, but the number of $G_+$ quanta is always the 
same as the number of $G_-$ quanta, and similarly for~$H$.
 The analog of (\ref{vac-tempk}) for standing waves is 
 the tensor product of (\ref{vac-temp}) for the cosine mode with its 
partner for the sine mode; 
 in that case, all occupation numbers must 
be even, but there is no constraint relating those for  $c$ 
quanta to those for $s$ quanta. 

 Let $\rho_{n\tilde n,n'\tilde n'}$ be the matrix element of the 
statistical operator between the state $\<n\tilde n|$ with $n$ 
(physical) $\beta_+$ quanta and $\tilde n$ $\beta_-$ quanta
  and another such state $|n'\tilde n'\>$.
 A long calculation parallel to that leading to (\ref{density})
 yields
\begin{equation}
\rho_{n\tilde nn'\tilde n'}=0 \quad 
 \hbox{if $n-\tilde n \ne n'-\tilde n'$}, 
 \label{oddk}\end{equation}
 and otherwise
  \begin{eqnarray}
\rho_{n\tilde nn'\tilde n'}&=&
 \frac1{\cosh^2\Omega\,\cosh^2\Psi} \, 
 \sum_{m =0}^\infty \sum_{l=0}^p \sum_{l'=0}^{p'}
  {\sqrt{n!\,\tilde n!\,n'!\,\tilde n'!}\,m!\,\tilde m! \over
2^{p+p'-2m-2\tilde m}\,l!\,l'!\, (p-l)!\, (p'-l')!}
   \nonumber \\
 &&{}\times 
P(l,p,m)P(l,p,\tilde m) P(l',p',m) P(l',p',\tilde m) 
 \nonumber\\
 &&{}\times
  \tanh^{p+p'-l-l'}\Omega \, \tanh^{l+l'}\Psi, 
  \label{densityk}\\
 && \tilde m = n+m-\tilde n, \quad
 p = n+m, \quad p'=n'+m. \nonumber
\end{eqnarray}
A corollary of the momentum conservation constraint (\ref{oddk})
 is that $n+\tilde n$ has the same parity as $n'+\tilde n'$
 in any nonvanishing matrix element;
 this is necessary for consistency with~(\ref{odd}) and with the 
general principle that Bogolubov transformations ``create'' quanta 
only in pairs. 

 \section{Number observables}\label{sec:5}
 Henceforth it is convenient to write formulas in terms of
 \begin{equation}
 A = \tanh \theta \quad\hbox{and}\quad
 B = \tanh\phi = e^{-\epsilon/2T}.
 \label{AB}\end{equation}
Along with (\ref{params}) we have (noting that $|AB|<1$)
 \begin{eqnarray}
 \cosh\Omega = {1+AB\over \sqrt{(1-A^2)(1-B^2)}} \,,
 &\ & \sinh\Omega = {A+B \over \sqrt{(1-A^2)(1-B^2)}} \,, 
\nonumber\\
 \cosh\Psi   = {1-AB\over \sqrt{(1-A^2)(1-B^2)}} \,,
 &\ & \sinh\Psi   = {A-B \over \sqrt{(1-A^2)(1-B^2)}} \,.
\label{coshsinh}\end{eqnarray}

 Our formalism provides two ways to calculate the expectation value 
of an operator.
 One may use the density matrix, (\ref{density}) or 
(\ref{densityk}).
 Alternatively, one can express the operator in terms of the 
operators $G$ and $H$ and their adjoints and take its matrix 
element in the pure state vector (\ref{vac-temp}) or 
 (\ref{vac-tempk}).
 For some observables the second method is much easier and also 
displays the $\theta\leftrightarrow\phi$ (or $A\leftrightarrow B$) 
symmetry to the maximal extent. 

 Consider a fixed mode pair $\pm \z9k$.
Recall that all our creation and annihilation operators
  carry subscripts $c$, $s$, $+$, or $-$; 
 we routinely omit not only any reference to the vector $\z9k$
  but also the subscript 
 (especially $c$ or $+$), 
  when no ambiguity results.
When the sine mode is involved, the correct state vector is the 
product of (\ref{vac-temp}) with its sine partner, whose formula is 
the same as (\ref{vac-temp}) except for a factor $(-1)^{j+l}$ in 
the summand.
 For any mode type there is a number operator
 \begin{equation}
n = \beta^\dg \beta 
 = \frac1{\sqrt2} (G^\dg+H^\dg) \frac1{\sqrt2} (G+H) 
 = \frac12(G^\dg G + H^\dg H + G^\dg H + H^\dg G).
 \label{n-GH}\end{equation}

 \subsection{The mean number}\label{sec:5a}

 First consider $\<n_+\>$ or $\<n_-\>$.
   From (\ref{vac-tempk}) we see that only $G^\dg G$ and  $H^\dg H$
 contribute to $\<n\> = \{0,0,0,0|n|0,0,0,0\}$, because the other 
two terms in (\ref{n-GH}) destroy the equality of the $+$ and $-$ 
occupation numbers.
 Using the first of
 \begin{equation}
 \sum_{j=0}^\infty j x^j = {x\over (1-x)^2}\,, \quad
 \sum_{j=0}^\infty jx^j {(2j-1)!!\over (2j)!!} = {x\over2}
 (1-x)^{-3/2}, 
 \label{jseries}\end{equation}
along with (\ref{coshsinh}), one calculates 
 \begin{equation}
 \<n\> = {A^2+B^2 \over (1-A^2)(1-B^2)}\,.
\label{nexp}\end{equation}
 
 After chasing through several layers of definitions, one sees that 
 (\ref{nexp}) ought to coincide with (each term of) equation (72) of 
\onlinecite{KKS}, and it does.
 Incidentally, (72) is one of very few formulas in \onlinecite{KKS} that 
are (even implicitly) symmetric in $A$ and~$B$.
 But it is now clear why it, and 
 \emph{any expectation value that concerns only one traveling-wave mode},
 must have that symmetry:
The thermal Bogolubov transformation (\ref{Bogo-gen}) connecting that mode to one 
of the fictitious barred modes has the same algebraic form as the transformation 
(\ref{Bogo-Tk}) connecting that mode to another of the physical modes. 
When the expectation value is calculated, 
 \emph{all} other modes are effectively traced over, so there is nothing to
distinguish the fictitious partner mode, with mixing constant~$B$, from the
physical one, with mixing constant~$A$.

This argument does not apply to standing waves, because the 
transformation (\ref{Bogo-A})--(\ref{Bogo-sin})
does not have the same form as~(\ref{Bogo-Tk}); 
in fact, we shall soon see that the conclusion does not hold for such modes.
Nevertheless, a calculation based on 
(\ref{vac-temp}) shows that 
\begin{equation}
\<n_c\> = \<n_s\> = \<n\>.
\label{ncs}\end{equation} 
 In this case the second of formulas (\ref{jseries}) is used, and 
again the $G^\dg H$ and $H^\dg G$ terms in (\ref{n-GH}) do not 
contribute, but this time for a different reason:  they produce 
 occupation numbers of odd parity, which do not occur in 
 $\{0,0|$.

 \subsection{The second moment}\label{sec:5b}

The operator $n^2$ is
\begin{equation}
 n^2= {\textstyle \frac14} (O_R + O_P + O_I),
 \label{n2}
\end{equation}
where the terms
 \begin{equation}
 O_R = (G^\dg G)^2 +(H^\dg H)^2 + 4 G^\dg G H^\dg H + G^\dg G + 
H^\dg H
 \end{equation}
are \emph{relevant}, the terms
 \begin{equation}
 O_P = (G^\dg)^2 H^2 + (H^\dg)^2 G^2 
 \end{equation}
are \emph{partially relevant}, and the terms
\begin{equation}
 O_I = 2[   (G^\dg)^2 GH + (H^\dg)^2 HG+ G^\dg G^2 H^\dg
 + H^\dg H^2 G^\dg + G^\dg H + H^\dg G]
 \end{equation}
 are \emph{irrelevant} because they contribute nothing to 
$\<n^2\>$.
 The partially relevant terms conserve parity but not momentum, so 
they contribute in the context of (\ref{vac-temp}) but not 
(\ref{vac-tempk}).

 For $\<n_+^2\>$ or $\<n_-^2\>$ a calculation like that of $\<n\>$ 
leads to a formula that can be abbreviated as
 \begin{equation}
 \<n^2\> = 2 \<n\>^2 + \<n\>.
\label{n2exp} \end{equation}
 Since $\<(n-\<n\>)^2\> = \<n^2\> - \<n\>^2$,
 it follows that the second moment of the number distribution is
\begin{equation}
\<(n-\<n\>)^2\> = \<n\>^2 + \<n\>.
 \label{n2mom}\end{equation}
 Again, these quantities are symmetric in $A$ and $B$.

A parallel calculation yields 
\begin{equation} 
\<n_c^2\> =\<n_s^2\>=
\frac{(1+B^2)[A^4 + B^2 + 2 A^2 (1+B^2)]}{{(1-A^2)}^2{(1-B^2)}^2}\,. 
\label{n2cs}\end{equation}
Here the symmetry between $A$ and $B$ is destroyed by the contribution of the 
partially relevant terms.

 \subsection{The second cumulant}\label{sec:5c}

Kocharovsky et al.\ \cite{KKS} do not give a formula for $\<n^2\>$;
 the closest point of comparison is the \emph{generating cumulant}
 $\tilde\kappa_2\,$.
One case of their formula (71) is, after reexpression in terms of 
$A$ and $B$ rather than $z(A)$ and $z(-A)$,
 \begin{eqnarray}
\tilde\kappa_2 &=&
 {A^2B^4 +A^4+4A^2B^2 +B^4+A^2  \over (1-A^2)^2(1-B^2)^2}
   \label{kappa}\\
&=& \<n^2\> + {A^2(B^2+1)^2 \over (1-A^2)^2(1-B^2)^2}\,.
\nonumber\end{eqnarray}
 Note that the second term is not symmetric in $A$ and $B$.

 This cumulant refers to a \emph{pair} of modes, $\pm\z9k$.
 Its definition is 
 \begin{eqnarray}
\tilde\kappa_2 &=& 
 {\textstyle\frac12}
 \bigl[ \<(n_++n_- -2\<n\>)^2 - 2\<n\>\bigr] \nonumber\\
 &=& \<n^2\> -2\<n\>^2 - \<n\> + \<n_+n_-\> \nonumber\\
 &=& \<n_+n_-\>,
 \label{kapparel}\end{eqnarray}
where (\ref{n2exp}) has been used at the last step
 (and $\<n_\pm\>=\<n\>$ from the beginning).   
 Thus the asymmetry of $\tilde\kappa_2$ in the two parameters
 arises from ``interference'' between two modes.
 The quantity is readily calculated by our method:
 \begin{equation}
 n_+n_-= {\textstyle \frac14} (O_N + O_T + O_O),
 \label{nn}\end{equation}
 where
\begin{eqnarray}
 O_N &=& G^\dg G G_-^\dg G_- + H^\dg H H_-^\dg H_- 
+ G^\dg G H_-^\dg H_- + H^\dg H G_-^\dg G_- \,,
\nonumber\\
 O_T &=& G^\dg G_-^\dg HH_- + GG_- H^\dg H_-^\dg
 +G^\dg G_- H_-^\dg H + G_-^\dg G H^\dg H_-\,,
\end{eqnarray}
and $O_O$ comprises $8$ terms with either three $G$ operators and 
one $H$ or vice versa.
 It is easy to see that $O_O$ and the last two terms of $O_T$ do 
not contribute to the expectation value of $n_+n_-$ in 
$|0,0,0,0\}$.
 The calculation of $\frac14\<O_N\>$ 
 is very similar to that of $\<n^2\>$ and results in the symmetric 
expression
 \begin{equation}
 {A^4 + 4A^2B^2 + B^4 + \frac12(A^2+B^2)(1+A^2B^2) \over
(1-A^2)^2(1-B^2)^2}\,.
 \label{ON}\end{equation}
For $\frac14\<O_T\>$ one finally gets an asymmetric expression,
 \begin{equation}
  {\frac12(A^2-B^2)(1-A^2B^2)\over (1-A^2)^2(1-B^2)^2}\,.
 \label{OT}\end{equation}
 Adding (\ref{ON}) and (\ref{OT}) and rearranging, one reproduces
 (\ref{kappa}).

\subsection{The quadratic moments for standing waves} \label{sec:5d}

The analog of $\<n_+n_-\>$ for the standing modes is
$\<n_cn_s\>$, the expectation value being taken in $|0,0\}$ for the cosine and 
sine modes independently.
Thus
\begin{equation}
\<n_c n_s\> = \<n_c\> \<n_s\>
= \<n\>^2  = \left[\frac{A^2 + B^2}{(1-A^2)(1-B^2)}\right]^2.
\label{ncns}\end{equation}

 Clearly one must have
\begin{equation}
\<(n_c+n_s)^2\> = \< (n_+ + n_-)^2\>,
\label{totalsq}\end{equation}
whence by (\ref{n2exp}) and (\ref{ncns}) one has
\begin{eqnarray}
{\textstyle\half}[\<n_c^2\>+ \<n_s^2\>]&=&
 - \<n_c n_s\> + \<n^2\>+ \<n_+ n_-\> 
\nonumber\\
&=& \<n\>^2 +\<n\> + \<n_+ n_-\>,
\label{sumsq}\end{eqnarray}
which simplifies to the quantity  (\ref{n2cs})
upon substitution from (\ref{nexp}) and (\ref{kappa})--(\ref{kapparel}).
Given that $\<n_c^2\>= \<n_s^2\>$, and that $\<n_+n_-\>$ is known,
this argument is the easiest way to derive~(\ref{n2cs}).

\section{The generating function}\label{sec:6}
\subsection{The single-mode density matrix revisited}\label{sec:6a}

The matrix elements $\rho_{nn'}=\langle n|\rho|n'\rangle$ 
of (\ref{odd}) and (\ref{density}) have been
obtained by regarding the statistical operator $\rho$ of (\ref{dens-phys}) as
referring to one mode of a two-mode system that is in the pure state
$|0,0\}\{0,0|$.
A different, and somewhat more direct, method of calculating $\rho_{nn'}$
first constructs the generating function
\begin{equation}
  \label{eq:g-def}
  g(x,y)=\sum_{n,n'=0}^{\infty}
\frac{x^n\langle n|\rho|n'\rangle y^{n'}}{\sqrt{n!\,n'!\,}}
\end{equation}
and then expands it in powers of $x$ and $y$.
The bra and ket states that show up here as the summations over $n$ and $n'$
are recongnized as the well known coherent states (of Bargmann type), the
eigenbras of $\beta\adj$ and eigenkets of $\beta$,
\begin{equation}
  \label{eq:Bargman}
  \sum_{n=0}^{\infty}\frac{x^n\langle n|}{\sqrt{n!}}
  \bigl(\beta\adj-x\bigr)=0\,,
\qquad
\bigl(\beta-y\bigr)
\sum_{n'=0}^{\infty}\frac{|n'\rangle y^{n'}}{\sqrt{n'!}}=0\,.
\end{equation}

These eigenvector equations are exploited upon writing $\rho$ as a normally
ordered function of $\beta\adj$ and $\beta$; that is, all
$\beta\adj$s stand to the left of all $\beta$s.
We begin with
\begin{eqnarray}
  \label{eq:rho-normord}
  \rho&=&\bigl(1-e^{-\epsilon/T}\bigr)e^{-(\epsilon/T)b\adj b}=
         \bigl(1-e^{-\epsilon/T}\bigr)
         \exp\Bigl(-\bigl(1-e^{-\epsilon/T}\bigr)b\adj;b\Bigr)
\nonumber\\
&=&e^{\half({\beta\adj}^2-\beta^2)\theta}
         \bigl(1-e^{-\epsilon/T}\bigr)
         \exp\Bigl(-\bigl(1-e^{-\epsilon/T}\bigr)\beta\adj;\beta\Bigr)
   e^{-\half({\beta\adj}^2-\beta^2)\theta}\,,
\end{eqnarray}
where
\begin{equation}
  \label{eq:ordexp}
  e^{X;Y}=\sum_{n=0}^{\infty}\frac{X^nY^n}{n!}
\end{equation}
denotes the basic ordered exponential function, for which the much used
identity
\begin{equation}
  (1-\lambda)^{b\adj b}=e^{-\lambda b\adj;b}
\end{equation}
is a familiar application.

The $\beta\adj;\beta$-ordered version of $\rho$, 
\begin{equation}
  \label{eq:rho-ordered}
  \rho=\sqrt{\lambda^2-\mu^*\mu\,}\;
      e^{\half\mu{\beta\adj}^2}e^{-\lambda \beta\adj;\beta}
      e^{\half\mu^*\beta^2}
\end{equation}
with
\begin{eqnarray}
  \label{eq:lam+mu}
  \lambda&=&
  1-\frac{e^{-\epsilon/T}}{(\cosh\theta)^2-e^{-2\epsilon/T}(\sinh\theta)^2}\,,
\nonumber\\
\mu=\mu^*&=&\bigl(1-e^{-2\epsilon/T}\bigr)
\frac{\sinh\theta\,\cosh\theta}
{(\cosh\theta)^2-e^{-2\epsilon/T}(\sinh\theta)^2}\,,
\nonumber\\
\sqrt{\lambda^2-\mu^*\mu\,}&=&
\frac{1-e^{-\epsilon/T}}
{\sqrt{(\cosh\theta)^2-e^{-2\epsilon/T}(\sinh\theta)^2}}\,,
\end{eqnarray}
can now be found by a variety
of methods.
It suffices to verify that it
responds correctly to infinitesimal changes of $\theta$.

Now, when using (\ref{eq:rho-ordered}) in (\ref{eq:g-def}), the identities
(\ref{eq:Bargman}) permit the replacements $\beta\adj\to x$, $\beta\to y$,
and then the summation is elementary,
\begin{equation}
\sum_{n,n'=0}^{\infty}
\frac{x^n\langle n|n'\rangle y^{n'}}{\sqrt{n!\,n'!\,}}=e^{xy}\,.  
\end{equation}
We thus 
arrive at the explicit form of the generating function (\ref{eq:g-def}),
\begin{equation}
  \label{eq:g-expl}
  g(x,y)=\sqrt{\lambda^2-\mu^*\mu\,}\;
         e^{\half\mu x^2}e^{(1-\lambda)xy}e^{\half\mu^*y^2}\,.
\end{equation}
Such a two-dimensional Gaussian is a typical generating function for
linear systems, since all time-transformation functions are of this form if
the Heisenberg equations of motion are linear.
A recent example is the parametric oscillator investigated by Rashid and
Mahmood \onlinecite{RasMah},
who     combine the Maclaurin series of the three exponential
functions in (\ref{eq:g-expl})
and get an answer where one summation is still to be done, somewhat like the
situation in (\ref{density}) above.

Here we wish to put a different approach on record. 
It exploits the well known fundamental 
generating function for Bessel
functions of integer order (see 9.1.41 and 9.1.5 in \onlinecite{AbrSteg}),
\begin{equation}
  \label{eq:Bessel-gen}
e^{\half z(t-1/t)}
            =\sum_{k=-\infty}^{\infty}t^kJ_k(z)
            =\sum_{k=-\infty}^{\infty}t^k 
(-1)^{\half(|k|-k)}(z)\,,
\end{equation}
and one of the generating functions for Gegenbauer polynomials 
of index $k+\half$ (see 22.9.5 in \onlinecite{AbrSteg}), 
\begin{equation}
  \label{eq:Gegenb-gen}
  e^u\bigl( z/2\bigr)^{-k}J_k(z)=\sum_{l=0}^{\infty}
\frac{(2k)!}{k!\,(2k+l)!}\bigl(z^2+u^2\bigr)^{\half l}C_l^{(k+\half)}
\left(u\Big/\sqrt{z^2+u^2}\,\right)\,,
\end{equation}
valid for $k\geq0$.
Jointly they amount to
\begin{eqnarray}
  \label{eq:combined}
  e^{\half z(t-1/t)+u}&=&\sum_{k=-\infty}^{\infty}\sum_{l=0}^{\infty}
2^{-|k|}\bigl(zt\bigr)^{\half(|k|+k)}\bigl(-z/t\bigr)^{\half(|k|-k)}
\frac{(2|k|)!}{(|k|)!\,(2|k|+l)!}
\nonumber\\
&&\hphantom{\sum_{k=-\infty}^{\infty}\sum_{l=0}^{\infty}}\times
\bigl(z^2+u^2\bigr)^{\half l}
C_l^{(|k|+\half)}\left(u\Big/\sqrt{z^2+u^2}\,\right)\,.
\end{eqnarray}
The left-hand side thereof turns into the product of exponentials in
(\ref{eq:g-expl}) if one puts
\begin{equation}
  zt=\mu x^2\,,\quad -z/t=\mu^*y^2\,,\quad
u=(1-\lambda)xy\,,
\end{equation}
so that
\begin{equation}
  \bigl(z^2+u^2\bigr)^{\half}=\sqrt{(1-\lambda)^2-\mu^*\mu\,}\,xy\,,\qquad
  \frac{u}{\sqrt{z^2+u^2}}=\frac{1-\lambda}{\sqrt{(1-\lambda)^2-\mu^*\mu\,}}\,.
\end{equation}
Note that the latter, which is the argument of the Gegenbauer polynomial in
(\ref{eq:combined}), does not depend on $x$ or $y$. All dependence on $x$ and
$y$ is in the form of explicit powers.

Upon putting things together, we arrive at
\begin{eqnarray}
\frac{g(x,y)}{\sqrt{\lambda^2-\mu^*\mu}}&=&
 e^{\half\mu x^2+(1-\lambda)xy+\half\mu^*y^2}
\nonumber\\&=&
  \sum_{k=-\infty}^{\infty}\sum_{l=0}^{\infty}
2^{-|k|}\left(\mu x^2\right)^{\half(|k|+k)}
\left(\mu^*y^2\right)^{\half(|k|-k)}
\nonumber\\&&\hphantom{\sum_{k=-\infty}^{\infty}\sum_{l=0}^{\infty}}\times
\frac{(2|k|)!}{(|k|)!\,(2|k|+l)!}
\left(\sqrt{(1-\lambda)^2-\mu^*\mu\,}\,xy\right)^l
\nonumber\\&&\hphantom{\sum_{k=-\infty}^{\infty}\sum_{l=0}^{\infty}}\times
C_l^{(|k|+\half)}\left(\frac{1-\lambda}{\sqrt{(1-\lambda)^2-\mu^*\mu\,}}\right)
\,,
\label{eq:g-detail}
\end{eqnarray}
where $n=|k|+k+l$ is the total power of $x$ and $n'=|k|-k+l$ is that of $y$,
their difference $n-n'=2k$ and sum $n+n'=2|k|+2l$ being even.
We write
\begin{equation}
  k=\frac{n-n'}{2}\,,\quad l=n_<=\min\{n,n'\}\,,\quad 2|k|+l=n_>=\max\{n,n'\}
\end{equation}
to present (\ref{eq:g-detail}) as a sum over $n$ and $n'$.
This enables us to
identify the matrix element $\langle n|\rho|n'\rangle=\rho_{nn'}\,$, with the
outcome
\begin{eqnarray}
  \label{eq:rhonn'-expl}
  \rho_{nn'}&=&
\frac{(n_>-n_<)!}{\bigl(\half(n_>-n_<)\bigr)!}
\left(\frac{n_<!}{n_>!}\right)^{\half}
\sqrt{\lambda^2-\mu^*\mu}\,
\nonumber\\&&{}\times
\bigl(\mu/2\bigr)^{\half(n-n_<)}\bigl(\mu^*/2\bigr)^{\half(n'-n_<)}
\left((1-\lambda)^2-\mu^*\mu\right)^{\half n_<}
\nonumber\\&&{}\times
C_{n_<}^{\bigl(\half(n_>-n_<+1)\bigr)}
\left(\frac{1-\lambda}{\sqrt{(1-\lambda)^2-\mu^*\mu\,}}\right)
\end{eqnarray}
if $n-n'$ is even, and $\rho_{nn'}=0$ if $n-n'$ is odd.
In addition to the expressions of (\ref{eq:lam+mu}) we also meet here
\begin{equation}
  (1-\lambda)^2-\mu^*\mu
=\frac{e^{-2\epsilon/T}(\cosh\theta)^2-(\sinh\theta)^2}
      {(\cosh\theta)^2-e^{-2\epsilon/T}(\sinh\theta)^2}\,.
\end{equation}

In fact, the result (\ref{eq:rhonn'-expl}) 
is a bit too general for our purposes because we have the case
of $\mu=\mu^*$, which simplifies matters somewhat.
Expressed in terms of $A=\tanh\theta$ and $B=e^{-\epsilon/2T}$ we have, for
$n-n'$ even,
\begin{eqnarray}
  \label{eq:rhonn'-explAB}
  \rho_{nn'}&=&
\frac{(n_>-n_<)!}{\bigl(\half(n_>-n_<)\bigr)!}
\left(\frac{n_<!}{n_>!}\right)^{\half}
\left(\frac{1-A^2}{1-A^2B^4}\right)^{\half}(1-B^2)
\nonumber\\&&{}\times
\left(\half\frac{A(1-B^4)}{1-A^2B^4}\right)^{\half(n_>-n_<)}
\left(\frac{B^4-A^2}{1-A^2B^4}\right)^{\half n_<}
\nonumber\\&&{}\times
C_{n_<}^{\bigl(\half(n_>-n_<+1)\bigr)}
\left(\frac{(1-A^2)B^2}{\sqrt{(1-A^2B^4)(B^4-A^2)}}\right)\,.
\end{eqnarray}

\subsection{Diagonal terms}\label{sec:6b}
The Gegenbauer polynomials of index $\half$ are the Legendre 
polynomials, \penalty-8000 
$C_n^{(\half)}=P_n\,$, so that the diagonal matrix elements are given by
\begin{equation}
  \label{eq:diag1}
  \langle n|\rho|n\rangle=\rho_{nn}=\sqrt{\lambda^2-\mu^*\mu}
\left((1-\lambda)^2-\mu^*\mu\right)^{\half n}
P_n\left(\frac{1-\lambda}{\sqrt{(1-\lambda)^2-\mu^*\mu\,}}\right).
\end{equation}
Upon recalling the familiar generating function for Legendre polynomials,
\begin{equation}
  \sum_{n=0}^{\infty}t^nP_n(x)=\frac{1}{\sqrt{1-2tx+t^2}}\,,
\end{equation}
we obtain the corresponding generating function for $\rho_{nn}\,$,
\begin{equation}
  \label{eq:diag2}
  \sum_{n=0}^{\infty}e^{inu}\rho_{nn}=
\left[\frac{\lambda^2-\mu^*\mu}
{1-2(1-\lambda)e^{iu}+\bigl[(1-\lambda)^2-\mu^*\mu]e^{2iu}}\right]^{\half},
\end{equation}
which one recognizes to be identical with (26) in \onlinecite{EFP} when
the differences in notation are taken into account.

 \subsection{Comparison}\label{sec:6c}
The dependence of (\ref{eq:rhonn'-explAB}) on $A$ and $B$, although complicated, 
is nicely consistent with the observations in Sec.~\ref{sec:3}.
Only even powers of $B$ appear; this is the symmetry of
(\ref{density}) under the interchange of $\Omega$ and $\Psi$.
Similarly, the behavior under change of sign of $\Omega$ and $\Psi$
is implemented by the factor $A^{(n_>-n_<)/2}$.

To establish direct contact between (\ref{eq:rhonn'-explAB}) and  
(\ref{density}) we would need 
to expand
\begin{equation}
 \rho_{nn'}\cosh\Omega\cosh\Psi=\rho_{nn'}\,\frac{1-A^2B^2}{(1-A^2)(1-B^2)} 
\end{equation}
in powers of
\begin{equation}
  \tanh\Omega=\frac{A+B}{1+AB}\quad\mbox{and}\quad\tanh\Psi=\frac{A-B}{1-AB}\,,
\end{equation}
which looks like an unfairly difficult homework exercise.
We have evaluated  (\ref{eq:rhonn'-explAB}) and  
(\ref{density})  numerically for a variety of values of the parameters,
always finding excellent agreement.

Thus the method of the generating function has provided a formula, 
(\ref{eq:rhonn'-explAB}), that is simpler than the one provided by the method of 
the thermal Bogolubov transformation, (\ref{density}).
(Numerically, the latter involves more computation and requires truncation of an 
infinite series.)
Nevertheless, the thermal Bogolubov method has given us an interesting 
and elegant way of looking at the problem,
which illuminates its symmetries and  may lead to further insights in the 
future.
  Furthermore, for observables of 
the sort studied in Sec.~\ref{sec:5}
the calculations based on the thermal pure state vectors, (\ref{vac-temp}) or 
 (\ref{vac-tempk}), appear to be 
competitive with the conventional methods. 

 \section*{Acknowledgments}
The authors acknowledge  support from
Air Force Research Laboratories (Rome, New York), DARPA-QuIST, TAMU
Telecommunication and Informatics Task Force (TITF) initiative, the
Office of Naval Research, and the Welch Foundation. 
BGE and MDP wish to thank Herbert Walther for the hospitality and support
they enjoyed at the MPI f\"ur Quantenoptik where part of this work was done.


\begin{thebibliography}{99}\frenchspacing

\bibitem{Bek} J. D. Bekenstein,
{\sl Phys. Rev. D \bf 7}, 2333 (1973); {\bf 12}, 3077 (1975).

\bibitem{BekM} J. D. Bekenstein and A. Meisels, 
{\sl Phys. Rev. D \bf15}, 2775 (1977).

\bibitem{Haw} S. W. Hawking, {\sl Commun. Math. Phys. \bf43}, 199 (1975);
{\sl Phys. Rev. D \bf13}, 191 (1976);
{\bf 14}, 2460 (1976).

\bibitem{otherBH} B. S. DeWitt, {\sl Phys. Reports\/ \bf19}, 295  (1975);
R. M. Wald, {\sl Commun. Math. Phys. \bf45}, 9 (1975);
{\sl Phys. Rev. D \bf13}, 3176 (1976);
L. Parker, {\sl Phys. Rev. D \bf12}, 1519 (1975).

\bibitem{Un} W. G. Unruh, {\sl Phys. Rev. D \bf14}, 870 (1976).

\bibitem{HH} J. B. Hartle and S. W. Hawking, {\sl Phys. Rev. D \bf13},
2188 (1976).

\bibitem{Ful} S. A. Fulling, {\sl J. Phys. A \bf 10}, 917 (1977).

\bibitem{GP} G. W. Gibbons and M. J. Perry, 
{\sl Phys. Rev. Lett. \bf 36}, 985 (1976);
{\sl Proc. Roy. Soc. A \bf 358}, 467 (1978).

\bibitem{Sew} G. L. Sewell, {\sl Ann. Phys. \rm(N.Y.) \bf 141}, 201 (1982).

\bibitem{revs} S. A. Fulling and S. N. M. Ruijsenaars,
{\sl Phys. Reports\/ \bf 152}, 135 (1987);
 B. S. Kay and R. M. Wald, {\sl Phys. Reports\/ \bf 207}, 49 (1991).

\bibitem{KMS} R. Kubo, {\sl J. Phys. Soc. Japan\/ \bf12}, 570 (1957);
P. C. Martin and J. Schwinger, {\sl Phys. Rev. \bf 115}, 1342 (1959).

\bibitem{HHW} R. Haag, N. M. Hugenholtz, and M. Winnink,
{\sl Commun. Math. Phys. \bf 5}, 215 (1967).

\bibitem{Par} L. Parker, {\sl Nature\/ \bf261}, 20 (1976).

\bibitem{Is} W. Israel, {\sl Phys. Lett. A \bf57}, 107 (1976).

\bibitem{AW} H. Araki and E. J. Woods, 
{\sl J. Math. Phys. \bf 4}, 637 (1963).

\bibitem{TU}
Y. Takahashi and H. Umezawa, {\sl Collective Phenom. \bf 2}, 55 (1975);
H. Umezawa, H. Matsumoto, and M. Tachiki,
{\sl Thermo Field Dynamics and Condensed States\/}
(North-Holland, Amsterdam, 1982).

\bibitem{BW} J. J. Bisognano and E. H. Wichmann, 
{\sl J. Math. Phys. \bf16}, 985 (1975); {\bf17}, 303 (1976).

\bibitem{GH} G. W. Gibbons and S. W. Hawking,
{\sl  Phys. Rev. D \bf15}, 2738 (1977).


\bibitem{B} N. N. Bogolubov, {\sl J. Phys. \rm(USSR) \bf11}, 23 (1947).

\bibitem{Bbook} N. N. Bogoliubov, {\sl Lectures on Quantum Statistics},
Vol. 1 (Gordon and Breach, New York, 1967) 
(translation of Ukrainian original, 1949).


\bibitem{ZB} V. A. Zagrebnov and J.-B. Bru, {\sl Phys. Reports\/ \bf350},
291 (2001).


\bibitem{KKS} V. V. Kocharovsky, Vl. V. Kocharovsky, and M. O. Scully,
{\sl Phys. Rev. A \bf61}, 053606 (2000).

\bibitem{expt}  E. A. Cornell and C. E. Wieman, Nobel lecture,
{\sl Rev. Mod. Phys. \bf74}, in press (July, 2002);
J. R. Anglin and W. Ketterle, {\sl Nature\/ \bf416}, 211 (2002).

\bibitem{expt-amps} 
A. Brunello, F. Dalfovo, L. Pitaevski, and S.
Stringari, {\sl Phys. Rev. Lett. \bf85}, 4422 (2000);
J. M. Vogels, K. Xu, C. Raman, J. R. Abo-Shaeer, and W.
Ketterle, {\sl Phys. Rev. Lett. \bf88}, 060402 (2002).


\bibitem{EFP}  B.-G. Englert, S. A. Fulling, and M. D. Pilloff,
{\sl Optics Commun.}, in press ({\tt quant-ph/0205023}).

\bibitem{Knu} D. E. Knuth, {\sl The Art of Computer Programming},
Vol.~1, 2nd ed. (Addison--Wesley, Reading, 1973), Sec.~1.2.6.

\bibitem{GR} I. S. Gradshteyn and I. M. Ryzhik,
{\sl Table of Integrals, Series, and Products\/} 
(Academic Press, New York, 1980),
Sec.~8.96.

\bibitem{PWZ} M. Petkov\v{s}ek, H. S. Wilf, and D. Zeilberger, {\sl A = B\/} 
(Peters, Wellesley, 1996).

\bibitem{RasMah}
M. A. Rashid and A. Mahmood,
\textsl{J. Phys.\ A: Math.~Gen.\ } \textbf{34}, 8185 (2001).

\bibitem{AbrSteg}
M. Abramowitz and I. A. Stegun, eds.,
\textsl{Handbook of Mathematical Functions\/} (Dover, New York, 1972).

\end{thebibliography}
\end{document}